\documentclass[aps,twocolumn,epsf]{revtex4}
\usepackage{graphics}
\usepackage{graphicx}
\usepackage{epsfig}

\begin{document}
\input epsf.tex
\bibliographystyle{apsrev}

\title{Alternative derivation of the response of interferometric gravitational wave detectors}
\author{Neil J. Cornish}
\affiliation{Department of Physics, Montana State University, Bozeman, MT 59717}

\begin{abstract}
It has recently been pointed out by Finn that the long-standing derivation of the response
of an interferometric gravitational wave detector contains several errors.
Here I point out that a contemporaneous derivation of the gravitational wave response for
spacecraft doppler tracking and pulsar timing avoids these pitfalls, and when adapted to
describe interferometers, recovers a simplified version of Finn's derivation.
This simplified derivation may be useful for pedagogical purposes.
\end{abstract}
\pacs{}

\maketitle

Finn~\cite{Finn:2008np} has shown that the standard
derivation~\cite{Weiss:1972,Forward:1978zm}, of the
the response of an interferometric gravitational wave detector is based on several
unjustified assumptions. These errors propagated undetected through the literature for
over 35 years. The main problem with the standard derivation is that it neglects the
lensing of the photon path by the the gravitational waves. The errors are easy to miss
when working in the transverse-traceless gauge, where the tying of the coordinates to free
test particles has the effect of absorbing the wave motion into the coordinate
system. Luckily, fortuitous cancellations in the transverse-traceless gauge lead to the correct
final result, so that the response formalism developed for gravitational wave detectors
such as LIGO~\cite{Saulson:1994} and LISA~\cite{Cornish:2002rt,Rubbo:2003ap} are unaffected.

The purpose of this comment is to point out that not all of the original derivations
of the gravitational wave response are flawed, and that as early as 1975 Estabrook
and Wahlquist~\cite{Estabrook:1975a} published a derivation that is equivalent to
Finn's when applied to interferometric detectors. Their derivation considered
gravitational wave detection by spacecraft doppler tracking, but it has since
been generalized to describe the gravitational wave response of
pulsar timing arrays~\cite{Detweiler:1979wn} and the fractional frequency shifts
measured by the LISA observatory (see Ref.~\cite{Vallisneri:2004bn} and the
appendix of Ref.~\cite{Rubbo:2003ap}).

Estabrook and Wahlquist start by considering a weak
gravitational wave propagating in the $z$ direction, which in the transverse-traceless
gauge is described by the line element (in units where the speed of
light $c=1$)
\begin{equation}
ds^2 = -du dv + (1+h_{xx})dx^2 + (1+h_{yy})dy^2 + 2 h_{xy} dxdy \, .
\end{equation}
Here $h_{ij}(u)$, $h_{yy}=-h_{xx}$, $u=t-z$ and $v=t+z$. They recognized
that the metric is independent of the
coordinates $x,y,v$, so there exist three
Killing vectors $\vec\partial_x$, $\vec\partial_y$, $\vec\partial_v$. The doppler
shifts caused by the gravitational wave follow immediately from the constancy
of the photon momentum one-form components $p_x$, $p_y$, $p_v$ and the null condition
$p_\mu p^\mu = 0$.

This elegant derivation can easily be generalized to describe the response of a
laser interferometer. In the transverse traceless gauge the coordinate acceleration
vanishes, so free test particles stay fixed at the same spatial coordinates, and
the coordinate time $t$ corresponds to the proper time $\tau$ along the world-line
of a test particle. As shown in Figure 1 of Ref.~\cite{Finn:2008np}, we need to compute
the difference in light travel times along the two arms of the interferometer. The
outward and return journeys along each arm can all be computed in the same way, so
we only need consider a single pass, which, without loss of generality, can be taken
as the null geodesic connecting the events $\vec{x}_1 \rightarrow (0,0,0,0)$ and
$\vec{x}_2 \rightarrow (t,x,y,z)$. The null condition and the three Killing vectors
yield four equations for the components of photon's 4-velocity $s^\mu = d x^\mu/d\lambda$:
\begin{eqnarray}\label{eom}
&& -s^u s^v + (1+h_{xx}) (s^x)^2 + (1+h_{yy}) (s^y)^2 +2h_{xy} s^x s^y = 0 \nonumber \\
&& (1+h_{xx})s^x + h_{xy} s^y = \alpha_1 \nonumber \\
&& (1+h_{yy})s^y + h_{xy} s^x = \alpha_2 \nonumber \\
&& s^u = -2 \alpha_3 \, .
\end{eqnarray}
Here the $\alpha_i$ are constants of integration that determine the photon path.
Using the $\alpha_1$ and $\alpha_2$ equations, the null condition can be re-written:
\begin{equation}
2\alpha_3 s^v + \alpha_1 s^x + \alpha_2 s^y = 0 \, .
\end{equation}
Since we are interested in describing weak gravitational waves $(h \ll 1)$, it is
permissible to solve these equations perturbatively such that
$x^\mu(\lambda) = x^\mu_0(\lambda) + \delta x^\mu(\lambda)$ and $\alpha_i = \alpha_i^0
+\delta\alpha_i$, where the zeroth order solution pertains when $h=0$. Simple
algebra yields
\begin{equation}
\alpha^0_1 = \frac{x}{\lambda_2-\lambda_1},
\quad \alpha^0_2 = \frac{y}{\lambda_2-\lambda_1},
\quad \alpha^0_3 = -\frac{L-z}{2(\lambda_2-\lambda_1)},
\end{equation}
and $t_0 = \sqrt{x^2 + y^2 + z^2} \equiv L$. As stressed by Finn~\cite{Finn:2008np}, the
photon path is lensed by the gravitational wave, and it is necessary to adjust ones
``aim'' (i.e. the $\alpha_i$) when $h\neq 0$. Using the fact that the coordinate location
of the test particles is unaffected by the gravitational wave, we have
$\delta{x}^i(\lambda_1)= \delta{x}^i(\lambda_2) = 0$, and expanding (\ref{eom})
to first order yields:
\begin{eqnarray}
\delta \alpha_1 &=& \frac{1}{(L-z)(\lambda_2-\lambda_1)}\left( x H_{xx} + y H_{xy}\right), \\
\delta \alpha_2 &=& \frac{1}{(L-z)(\lambda_2-\lambda_1)}\left( y H_{yy} + x H_{xy}\right), \\
\delta \alpha_3 &=& -\frac{\delta t}{2(\lambda_2-\lambda_1)} \, ,
\end{eqnarray}
where $H_{ij} = \int_{u_1}^{u_2} h_{ij}(u)\,  du$ and
\begin{eqnarray}
\delta t &=& \frac{1}{2 L (L-z)}\left( x^2 H_{xx} + y^2 H_{yy} + 2 xy H_{xy} \right)\nonumber \\
&=& \frac{1}{2} \frac{\hat{x}\otimes\hat{x} : {\bf H}}{1-\hat{k}\cdot\hat{x}} \, .
\end{eqnarray}
In the final expression for the time delay imparted by the gravitational wave, $\delta t$, I
have written the result in the usual coordinate independent form for a gravitational wave
propagating in the $\hat{k}$ direction~\cite{Cornish:2002rt}.

The derivation given above is equivalent to Finn's, the difference being that Finn started with
the second order differential equations for $x^\mu(\lambda)$ that follow from the geodesic equation,
while I started with the first integrals of the geodesic equation that follow from the presence
of the three Killing vectors. Writers of textbooks and review articles should pay attention
to Finn's article and avoid propagating the erroneous
derivation of the interferometer response function any further, and they may also want to consider
presenting the simpler ``Estabrook and Wahlquist style'' derivation outlined above.

\end{document}